\documentclass[aps,twocolumn,paper,showpacs,superscriptaddress, floatfix, longbibliography]{revtex4-2}
\usepackage[utf8]{inputenc}
\usepackage{amsmath,amssymb,physics}
\usepackage[hypertexnames=false]{hyperref}
\usepackage{datetime}
\usepackage{enumitem}
\usepackage{graphicx}
\usepackage{braket}
\usepackage{esdiff}
\usepackage{MnSymbol}
\usepackage{siunitx}
\usepackage{outlines}
\usepackage[normalem]{ulem}
\usepackage[dvipsnames]{xcolor}
\usepackage{xfrac}
\renewcommand{\selectlanguage}[1]{} 

\hypersetup{
    colorlinks,
    linkcolor={red!50!black},
    citecolor={blue!50!black},
    urlcolor={blue!80!black}
}

\begin{document}

\title{Universal entropy transport far from equilibrium across the BCS-BEC crossover}

\author{Jeffrey Mohan}
\author{Philipp Fabritius}
\author{Mohsen Talebi}
\author{Simon Wili}
\author{Meng-Zi Huang}
\email{mhuang@phys.ethz.ch}
\author{Tilman Esslinger}
\affiliation{Institute for Quantum Electronics \& Quantum Center, ETH Zurich, 8093 Zurich, Switzerland}
\date{\today}

\begin{abstract}
The transport properties of strongly interacting fermionic systems can reveal exotic states of matter, but experiments and theory have predominantly focused on bulk systems in the hydrodynamic limit describable with linear response coefficients such as electrical and thermal conductivity. In a ballistic channel connecting two superfluid reservoirs, recent experiments revealed a far-from-equilibrium regime beyond linear hydrodynamics where particle and entropy currents respond nonlinearly to biases of chemical potential and temperature, and their ratio is robust to the channel geometry. However, the origin of this robustness and its relation to the strong interparticle interactions remain unknown. Here, we study the coupled transport of particles and entropy tuning the interaction across the Bardeen-Cooper-Schrieffer to Bose-Einstein condensate (BCS-BEC) crossover, the reservoir degeneracy across the superfluid phase transition, as well as the local potentials and confinement of the channel. Surprisingly, the entropy advectively transported per particle depends only on the interactions and reservoir degeneracy and not on the details of the channel, suggesting that this property has its origin in the universal equilibrium properties of the reservoirs. In contrast, the magnitudes of the advective and diffusive entropy currents vary significantly with the channel details. The advective current increases monotonically towards the BEC side, which can be largely explained by the estimated superfluid gap in the channel. The Wiedemann-Franz law that links the advective and diffusive currents in Fermi liquids is most egregiously violated at unitarity, suggesting a change in the nature of the excitations responsible for entropy diffusion near unitarity. These observations pose fundamental questions regarding transport phenomena in strongly interacting Fermi systems far from equilibrium.
\end{abstract}

\maketitle

\section{Introduction}
Non-equilibrium quantum many-body systems are notoriously challenging to understand since there are few universal governing principles to apply beyond the often intractable microscopic equations of motion~\cite{polkovnikov_colloquium_2011, eisert_quantum_2015}. However, some such principles have emerged and been investigated in recent years~\cite{sieberer_universality_2023}, for example non-equilibrium equations of state~\cite{dogra_universal_2023}, self-organized criticality~\cite{helmrich_signatures_2020}, prethermal dynamics associated with non-thermal fixed points~\cite{prufer_observation_2018, erne_universal_2018, glidden_bidirectional_2021, garcia-orozco_universal_2022, gazo_universal_2023}, universal dynamics arising from scale invariance and other symmetries~\cite{zwerger_bcs-bec_2012, makotyn_universal_2014, eigen_universal_2018, saint-jalm_dynamical_2019, wang_exploring_2023, huh_universality_2024}, universal scaling laws in turbulent systems~\cite{navon_emergence_2016, hernandez-rajkov_universality_2023}, Kibble-Zurek dynamics~\cite{navon_critical_2015, clark_universal_2016, ko_kibblezurek_2019, lee_observation_2023}, and dynamical phase transitions~\cite{marino_dynamical_2022} to name a few. Another notable example is the universality observed in the linear response coefficients of many strongly correlated fermionic systems such as their linear-in-temperature resistivity~\cite{brown_bad_2019, xu_bad-metal_2019, phillips_stranger_2022}. While a generally accepted microscopic explanation for this universality does not yet exist, conjectures have been proposed that it originates from holographic duality~\cite{adams_strongly_2012, liu_quantum_2020, chowdhury_sachdev-ye-kitaev_2022, hartnoll_colloquium_2022}. Recently, it was shown theoretically that coupled particle and entropy transport, i.e. thermoelectricity, in bulk~\cite{davison_thermoelectric_2017} and more complex geometries~\cite{kruchkov_thermoelectric_2020} can be used to probe the many-body spectrum of such systems.

The BCS-BEC crossover~\cite{zwerger_bcs-bec_2012, strinati_bcsbec_2018} is a prototypical example of a strongly correlated fermionic system wherein no well-defined quasiparticles exist near the superfluid transition~\cite{sagi_breakdown_2015, mukherjee_spectral_2019}. The hydrodynamic transport properties of this system, being rooted in its universal equilibrium thermodynamics, have been predicted and experimentally confirmed to exhibit universal behavior~\cite{joseph_measurement_2007, cao_universal_2011, elliott_anomalous_2014, enss_universal_2019, patel_universal_2020, wang_hydrodynamic_2022, yan_thermography_2024, li_second_2022}. We have recently observed transport phenomena beyond the regime of linear hydrodynamics~\cite{fabritius_irreversible_2023}, wherein currents of particles and entropy flowing between two unitary fermi superfluids strongly coupled by a ballistic channel respond nonlinearly to biases in chemical potential and temperature. The measured entropy transported per particle was surprisingly robust to varying channel confinement, but the root of this robustness remains elusive. Here, we investigate how the characteristics of entropy transport are determined by global and local properties of the system. The former are the interaction strength and degeneracy that characterize the universal thermodynamics of the reservoirs, and the latter are the transverse confinement and local potential in the channel. Our observations suggest a possibly new type of far-from-equilibrium universality: the average entropy advectively transported per particle is determined only by the universal equilibrium thermodynamics of the reservoirs and not by the details of the channel.

\begin{figure}
    \centering
    \includegraphics{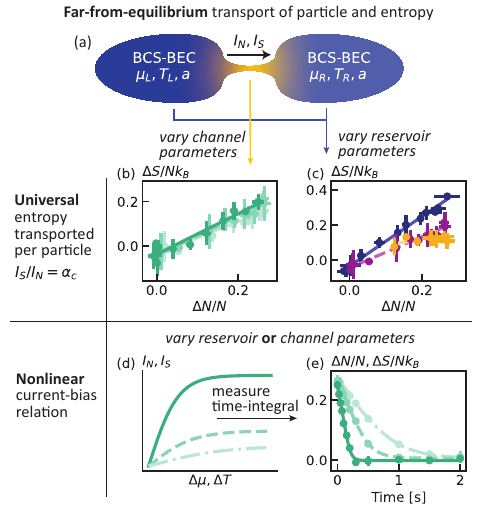} 
    \caption{\textbf{Far-from-equilibrium entropy transport in a strongly interacting Fermi gas showing universal dependence on the reservoirs' thermodynamics.} (a) Two reservoirs of fermionic \textsuperscript{6}Li exchange particles $N$ and entropy $S$ via currents $I_N$ and $I_S$ flowing through a ballistic channel wherein the gas is far from equilibrium. The nonlinear response of the currents to the biases of chemical potential $\Delta\mu$ and temperature $\Delta T$ (d,e) depends on parameters of the reservoirs and channel. The example data in (e) show $\Delta N/N$ with varying potential energy depth in the channel (darker green corresponds to higher $|V_g|$; see text). However, the entropy advectively transported per particle $\alpha_c = I_S/I_N = \dv*{\Delta S}{\Delta N}$ does not depend on the details of the channel [(b); same data as in (e)] and only varies with the universal thermodynamic properties of the reservoirs [(c); in these example data, interaction is varied from BEC (blue) to BCS (orange)]. The lines in (e) are integrated equations of motion from Eq.~\ref{eq:analytic_nonlinear} with fitted phenomenological coefficients. Data points and error bars are means and standard deviations of 3-5 repetitions.}
    \label{fig:fig1}
\end{figure}

\section{Universal transport far from equilibrium}
Fig.~\ref{fig:fig1} schematically illustrates the concept and central findings of the experiment. We prepare a degenerate gas of fermionic \textsuperscript{6}Li atoms separated into two harmonically-trapped reservoirs -- left $L$ and right $R$ -- and connected by a tunable channel formed by two intersecting $\mathrm{TEM}_{01}$-like beams. The two reservoirs are initially imbalanced in their atom number $\Delta N=N_L-N_R$ and entropy $\Delta S=S_L-S_R$ which induce biases in chemical potential $\Delta\mu=\mu_L-\mu_R$ and temperature $\Delta T=T_L-T_R$. Upon opening the channel, these biases drive the currents $I_N = -(1/2)\dv*{\Delta N}{t}$ and $I_S = -(1/2)\dv*{\Delta S}{t}$, which define the coupled equations of motion for the relaxation dynamics of $\Delta N(t)$ and $\Delta S(t)$ towards equilibrium $\Delta N = \Delta S = 0$.

To investigate the robustness of entropy transport to channel details and explore its dependencies, we explore the full range of available experimental parameters, varying properties of either the reservoirs or the channel. We map out the BCS-BEC crossover and normal-superfluid phase transition of the reservoirs by varying their dimensionless interaction strength $1/k_F a$ and total entropy per particle $S/N k_B = (S_L+S_R)/(N_L+N_R)k_B$; $a$ is the $s$-wave scattering length, $k_F$ is the trap-averaged Fermi wave vector, and $k_B$ is the Boltzmann constant. We map out the 1D-2D dimensional crossover of the channel with a variable horizontal transverse confinement frequency $\nu_x$ at the center of the channel and a tight vertical confinement $\nu_z=\SI{10.2(1)}{kHz}\gtrsim 5k_BT$. We also scan a large range of local potential in and around the channel with an attractive gate potential $V_g$ (App.~\ref{app:experimental_setup_and_sequence}). After repeatedly preparing the reservoirs in the same initial state $\Delta N(0), \Delta S(0)$ with a chosen set of parameters $1/k_Fa$, $S/Nk_B$, $\nu_x$, and $V_g$, we allow transport for varying time $t$, after which we adiabatically ramp down the channel beams and the interactions back to unitarity $1/k_Fa=0$ and finally acquire an absorption image of the reservoirs in the harmonic trap. From these images, we extract $\Delta N(t)$ and $\Delta S(t)$ using the known equation of state~\cite{ku_revealing_2012} and standard image analysis techniques~\cite{fabritius_irreversible_2023}. See App.~\ref{app:experimental_setup_and_sequence} for details of the experimental setup and procedure.

In previous work where we fixed $1/k_Fa=0$, $S/Nk_B$, $V_g$ and varied $\nu_x$~\cite{fabritius_irreversible_2023}, we found that the observed currents were reproduced by the nonlinear phenomenological model
\begin{align}\begin{split}
    \label{eq:phenomenological_model}
    I_N &= I_\mathrm{exc} \tanh\pqty{\frac{\Delta\mu + \alpha_c \Delta T}{\sigma}} \\
    I_S &= \alpha_c I_N + G_T \Delta T/T \\
    \pmqty{\Delta N \\ \Delta S} &= \frac{\kappa}{2} \pmqty{1 & \alpha_r \\ \alpha_r & \ell_r + \alpha_r^2} \pmqty{\Delta\mu \\ \Delta T}.
\end{split}\end{align}
This model possesses two modes of entropy transport: a nonlinear advective mode $I_S^a = \alpha_c I_N$, wherein each transported particle carries on average an entropy $\alpha_c$ with it, and a linear diffusive mode $I_S^d = G_T \Delta T/T$, which enables entropy transport without net particle transport. The advective mode is characterized by the Seebeck coefficient $\alpha_c$, excess current $I_\mathrm{exc}$, and nonlinearity coefficient $\sigma$. The diffusive mode is characterized by the thermal conductance $G_T$, and its strength relative to the advective mode is quantified by the Lorenz number $L=G_T/TG$, where $G=I_\mathrm{exc}/\sigma$ is the particle conductance in the linear-response limit at small bias or large $\sigma$. The compressibility $\kappa$, dilatation coefficient $\alpha_r$, and analogue Lorenz number $\ell_r$ are thermodynamic properties of the reservoirs~\cite{grenier_thermoelectric_2016} which we estimate here from the calculated equation of state of the BCS-BEC crossover~\cite{haussmann_thermodynamics_2007, haussmann_thermodynamics_2008}; see App.~\ref{app:thermodynamics} for details.

In the majority of conditions explored in this work, we find that the advective mode is nonlinear and dominates the diffusive mode. Its current-bias characteristics are sketched in Fig.~\ref{fig:fig1}(d) and correspond to the measured non-exponential relaxation shown in (e). Upon varying any parameter of the reservoirs ($1/k_Fa, S/Nk_B$) or channel ($\nu_x, V_g$), we observe a change in the relaxation dynamics [varying $V_g$ for the example data in (e); from light to dark green circles, $-V_g/k_B = 1.62(7),\,1.80(8),\,\SI{2.2(1)}{\micro K}$]. Despite this general variation of $I_N$ and $I_S$, we find by re-plotting the same data in (b) that the average entropy advectively transported per particle $\alpha_c = I_S/I_N = \dv*{\Delta S}{\Delta N}$ remains constant when varying any of the channel parameters. This is clear from the collapse of $\Delta S(t)$ vs.~$\Delta N(t)$ for all values of $V_g$. We observe in (c) that $\alpha_c$ only varies with the reservoir parameters that characterize their universal thermodynamics [varying $1/k_Fa$ for these example data; orange, purple, navy circles correspond to $1/k_Fa =-1.42(3),\,0,\,0.96(2)$ respectively]. Being a property of the far-from-equilibrium currents in the channel, $\alpha_c$ is \textit{a priori} expected to depend on every parameter of the system. However, our observation suggests that in this regime, $\alpha_c$ is a universal function of the equilibrium reservoirs properties $1/k_F a$ and $S/Nk_B$ analogously to thermodynamic properties like the particle number~\cite{zwerger_bcs-bec_2012}.

\begin{figure}
    \centering
    \includegraphics[width=\hsize]{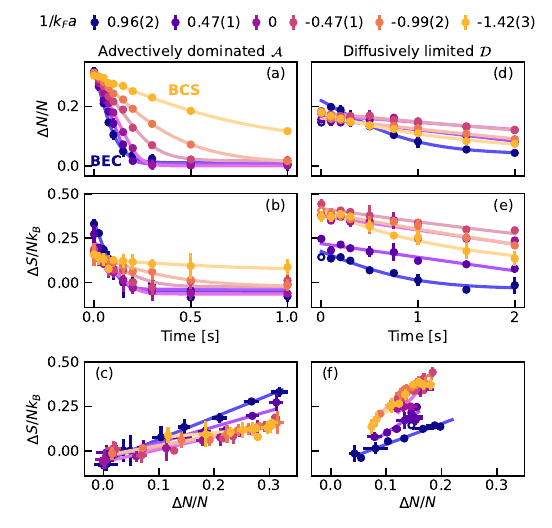}
    \caption{\textbf{System dynamics selectively probed in two eigenmodes vary strongly across the BCS-BEC crossover.} (a-c) At various interaction strengths $1/k_Fa$, we observe the evolution in particle and entropy imbalances starting from two complementary initial states with all other parameters kept constant, revealing distinct dynamics either dominated by nonlinear advection (a-c) or limited by linear diffusion (d-f). They are the two eigenmodes of the system's dynamics since they trace straight lines in the $(\Delta N, \Delta S)$ space (c,f), see text. The fits of Eq.~\ref{eq:analytic_nonlinear} (solid curves) yield the timescales for the two modes and the entropy advectively transported per particle [slopes in (c)]. The open markers in (d-f) are excluded from the fit since they have a significant overlap with the undesired advectively dominated eigenmode due to imperfect initial state preparation. Error bars represent standard deviations of 3-5 repetitions.}
    \label{fig:fig2}
\end{figure}

\section{Eigenmodes of transport}
In general, the advective and diffusive modes described by Eq.~\ref{eq:phenomenological_model} hybridize into two eigenmodes $\mathcal{E}$ of coupled particle and entropy transport when their timescales are similar at large Lorenz number $L$. Nevertheless, we identify an advectively dominated eigenmode $\mathcal{A}$ and a diffusively limited eigenmode $\mathcal{D}$ as advective transport is much faster than diffusive transport in most cases we explore; see App.~\ref{app:eigenmodes_of_transport} for details. We probe these two eigenmodes independently by preparing the reservoirs in two specific and complementary initial conditions $\Delta N(0), \Delta S(0)$. We ensure that only one eigenmode evolves at a time by verifying that the system traces a straight line through state space as it relaxes $\Delta S(t) = \Pi_\mathcal{E} \Delta N(t) + \Delta S_\mathrm{neq}$, where $\mathcal{E}=\mathcal{A}$ or $\mathcal{D}$. As explained in detail in App.~\ref{app:experimental_setup_and_sequence}, we prepare $\mathcal{A}$ primarily with a large particle number imbalance $\Delta N(0)$ and finely tune the entropy imbalance $\Delta S(0)$ to ensure a linear trajectory in state space. Likewise, we prepare $\mathcal{D}$ primarily with a large $\Delta S(0)$ and finely tune $\Delta N(0)$.

\begin{figure*}
    \centering
    \includegraphics{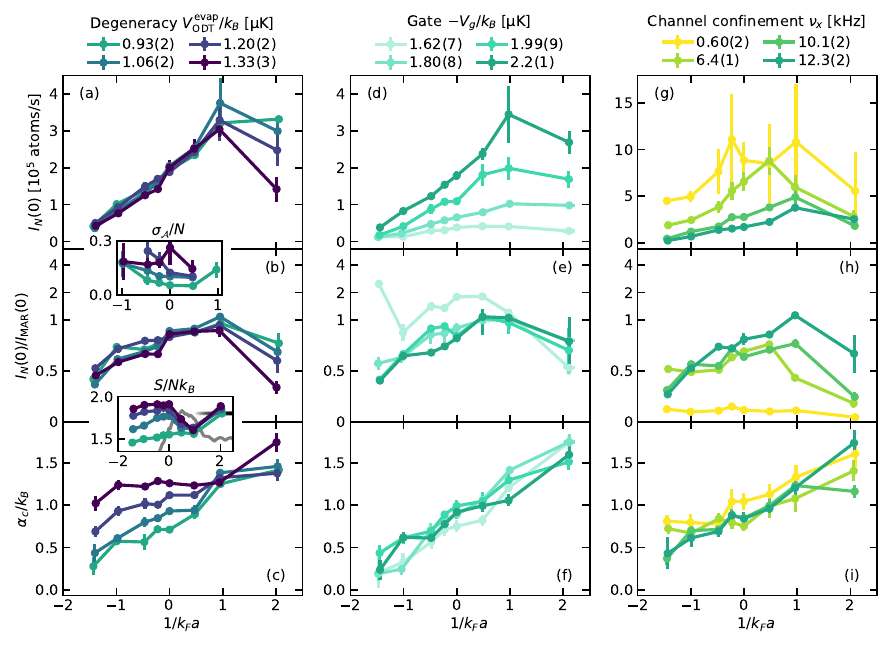}
    \caption{\textbf{Observation of universal advective particle and entropy transport across the BCS-BEC crossover with varying reservoir degeneracy, gate potential, and channel confinement}. The initial particle current $I_N(0)$ (a,d,g) weakly depends on the degeneracy of the reservoirs $S/Nk_B$ [inset between (b) and (c)] but strongly depends on the interactions $1/k_Fa$ and the channel properties $V_g$ and $\nu_x$. These dependencies are partially accounted for by expectations from multiple Andreev reflections $I_\mathrm{MAR}(0)$ [(b,e,h); the vertical axis is logarithmic above 1]. In contrast, the average entropy advectively transported per particle (Seebeck coefficient) $\alpha_c$ is independent of the channel properties [colors in (f,i)] and depends only on the universal thermodynamic properties of the reservoirs (c), $S/N k_B$ and $1/k_F a$. The solid gray curve plotted with $S/Nk_B$ is the estimated critical value at the superfluid transition in a harmonic trap (see App.~\ref{app:thermodynamics}) which is noisy due to numerical artifacts and deviates slightly from the known BEC limit $(S/Nk_B)_c \approx 1.801$~\cite{chen_thermodynamics_2005} (faded black line). This inset has the same vertical tick spacing as (c). The response is most nonlinear [inset between (a) and (b)] slightly on the BEC side and at low $S/Nk_B$. The nonlinearity $\sigma_\mathcal{A}$ is ill-defined in the linear response limit, so we only plot $\sigma_\mathcal{A}$ where transport is nonlinear (see App.~\ref{app:eigenmodes_of_transport}). The sea green data in all panels have $V_\mathrm{ODT}^\mathrm{evap}/k_B=\SI{0.93(2)}{\micro K}$, $-V_g/k_B=\SI{2.2(1)}{\micro K}$, and $\nu_x=\SI{12.3(2)}{kHz}$. Error bars are standard errors from least-squares fits.}
    \label{fig:fig3}
\end{figure*}

The advectively dominated and diffusively limited dynamics are shown in Fig.~\ref{fig:fig2}(a-c) and (d-f) respectively with varying $1/k_Fa$ and fixed $S/Nk_B=1.50(1)$, $\nu_x=\SI{12.3(2)}{kHz}$, $-V_g/k_B=\SI{2.2(1)}{\micro K}$. The uncertainties in $1/k_Fa$ are less than 2\% and are mainly due to atom number fluctuations that affect $k_F$. In both cases, the imbalances trace a straight line through state space as they relax towards equilibrium (c,f), ensuring that we are probing only one eigenmode. We observe that, even though the system parameters are identical in both cases, the relaxation times of $\mathcal{A}$ are significantly faster than $\mathcal{D}$ for all $1/k_F a$, and the slopes of the path in state space $\Pi_\mathcal{E}$ as well as their dependence on $1/k_F a$ differ between the two eigenmodes. Moreover, while the relaxation of $\mathcal{D}$ is always exponential $\propto e^{-t/\tau_\mathcal{D}}$, $\mathcal{A}$ is clearly non-exponential under most conditions, indicating nonlinear response of the advective mode.

Given the constraint that the path in state space is a line, Eq.~\ref{eq:phenomenological_model} has the analytic solution for the imbalances as a function of time in either eigenmode $\mathcal{E}$
\begin{align}
    \label{eq:analytic_nonlinear}
    \frac{\Delta N(t) - \Delta N_\mathcal{E}(\infty)}{\sigma_\mathcal{E}} &= \\
    & \hspace{-4em} \sinh^{-1}\qty{e^{-2I_\mathrm{exc}t/\sigma_\mathcal{E}} \sinh\bqty{\frac{\Delta N(0) - \Delta N_\mathcal{E}(\infty)}{\sigma_\mathcal{E}}}} \nonumber
\end{align}
where $\sigma_\mathcal{E}$ and $\Delta N_\mathcal{E}(\infty)$ are written in terms of the original coefficients of the model in App.~\ref{app:eigenmodes_of_transport}. The lines in Fig.~\ref{fig:fig1}(b,c,e) and Fig.~\ref{fig:fig2} are fits to Eq.~\ref{eq:analytic_nonlinear} with fitted $I_\mathrm{exc}$, $\sigma_\mathcal{E}$, $\Delta N_\mathcal{E}(\infty)$, $\Pi_\mathcal{E}$, and $\Delta S_\mathrm{neq}$. Eq.~\ref{eq:analytic_nonlinear} smoothly interpolates between the regimes of nonlinear ($\sigma \ll \abs{\Delta\mu + \alpha_c\Delta T}$) and linear ($\sigma \gg \abs{\Delta\mu + \alpha_c\Delta T}$) response. In the nonlinear limit, $\Delta N(t)$ is piecewise linear with initial current $I_N(0) = I_\mathrm{exc}$. In the linear limit, $\Delta N(t) \propto \Delta N(0) e^{-t/\tau_\mathcal{E}}$ with time constant $\tau_\mathcal{E}^{-1}=2I_\mathrm{exc}/\sigma_\mathcal{E}$ and initial current $I_N(0) = [\Delta N(0)-\Delta N_\mathcal{E}(\infty)]/2\tau_\mathcal{E}$. As shown in App.~\ref{app:eigenmodes_of_transport}, the slope of the advectively dominated eigenmode approximates the Seebeck coefficient $\Pi_\mathcal{A} \approx \alpha_c$ since advection is generally much faster than diffusion. The ratio of the advective and diffusive timescales, which provides a simple way to characterize the relative strength of the two modes, is proportional to the Lorenz number $\tau_\mathcal{A}/\tau_\mathcal{D} \approx L \ell_r/[\ell_r + (\alpha_c-\alpha_r)^2]^2$.

All data sets varying $1/k_Fa$ for each eigenmode (columns in Fig.~\ref{fig:fig2}) are prepared with the same initial conditions $\Delta N(0), \Delta S(0)$. The measured initial values with $1/k_Fa>0$ deviate from the values with $1/k_Fa\leq0$ due to inelastic collisions involving dimers that lead to particle loss and heating~\cite{inguscio_ultra-cold_2007, chin_feshbach_2010}. This is apparent even for $t=0$ because we hold the system at each $1/k_Fa$ for \SI{2.225}{s} before imaging regardless of the transport time $t$ to optimize thermal stability of the experimental apparatus. This introduces uncertainties in the measured $\Pi_\mathcal{E}$ deep in the BEC regime.

\section{Advectively dominated dynamics}
In Fig.~\ref{fig:fig3}, we study the characteristics of the advectively dominated eigenmode as functions of the interactions $1/k_Fa$. We also vary the global degeneracy, $S/Nk_B$, determined by the depth of the optical dipole trap at the end of evaporation $V_\mathrm{ODT}^\mathrm{evap}$ (a-c), and the channel properties $V_g$ (d-f) and $\nu_x$ (g-i). For the data varying reservoir degeneracy (a-c), the total entropy per particle $S/Nk_B$ measured after \SI{2.225}{s} in the trap without transport ($t=0$) is shown in the inset between (b) and (c) along with the calculated critical $S/Nk_B$ at the superfluid transition (gray curve). The increase in $S/N k_B$ deep on the BEC side is due to inelastic collisions. The cause of the attractor-like behavior at $1/k_Fa\approx1$ is unclear, but appears to be a balance of heating and cooling effects arising from inelastic collisions~\cite{dogra_can_2019}. The example data in Fig.~\ref{fig:fig1}(b,e) are the same as the data with $1/k_Fa=0$ and $-V_g/k_B = 1.62(7),\,1.80(8),\,\SI{2.2(1)}{\micro K}$ in Fig.~\ref{fig:fig3}(d-f); Fig.~\ref{fig:fig1}(c) corresponds to $1/k_Fa = -1.42(3),\,0,\,0.96(2)$ and $-V_g/k_B=\SI{1.80(8)}{\micro K}$ in Fig.~\ref{fig:fig3}(d-f); Fig.~\ref{fig:fig2}(a-c) corresponds to $V_\mathrm{ODT}^\mathrm{evap}/k_B = \SI{0.93(2)}{\micro K}$ in Fig.~\ref{fig:fig3}(a-c).

To compare the nonlinear and linear regimes on equal footing, we quantify their timescale with the initial particle current $I_N(0)$, which reduces to $I_\mathrm{exc}$ for non-exponential relaxation and $\Delta N(0)/2\tau_\mathcal{A}$ for exponential relaxation. The top row (a,d,g) shows that, for all parameters scanned, $I_N(0)$ increases monotonically as $1/k_F a$ is tuned from the BCS side $1/k_F a \approx -1.4$ to the BEC side $1/k_F a \approx 1$, after which it begins to drop likely due to inelastic losses. For all $1/k_F a$, $I_N(0)$ increases monotonically with increasing $\abs{V_g}$ and decreasing $\nu_x$, but is nearly independent of $V_\mathrm{ODT}^\mathrm{evap}$. The inset of (a), wherein we plot the re-scaled nonlinearity coefficient $\sigma_\mathcal{A}/N \propto \sigma$, shows that the advective response is most nonlinear ($\sigma_\mathcal{A}/N$ is smallest) at low $S/N k_B$ and slightly to the BEC side from unitarity. The nonlinearity has no clear dependence on $V_g$ or $\nu_x$ (not shown).

In previous work at unitarity~\cite{husmann_connecting_2015, huang_superfluid_2023, visuri_dc_2023, fabritius_irreversible_2023}, we found quantitative agreement between measurements of $I_N(0)$ and predictions from microscopic models based on multiple Andreev reflection (MAR). These models predict an initial current scale $I_\mathrm{MAR}(0) = n_m[16\Delta/3h + I_F(0)]$ where $\Delta$ is the equilibrium superfluid pairing gap at the contacts to the channel, $I_F(0) = 4E_F\Delta N(0)/3hN$ is the initial current of a non-interacting Fermi gas at zero temperature with a single transmission mode, $n_m$ is the equilibrium number of occupied transmission modes in the channel, and $h$ is Planck's constant; see App.~\ref{app:natural_scales_of_transport} for details. Normalizing $I_N(0)$ to $I_\mathrm{MAR}(0)$ (b,e,h) causes $I_N(0)$ to collapse remarkably well, but leaves significant residual dependence of $I_N(0)/I_\mathrm{MAR}(0)$ on $1/k_Fa$. Indeed, the mean-field picture of MAR is only valid in the BCS limit as it breaks down beyond the splitting point on the BEC side and produces qualitatively different current-bias characteristics than we observe~\cite{setiawan_analytic_2022}. The data with the smallest $\abs{V_g}$ and $\nu_x$ do not collapse as well, though these two conditions have large uncertainties in the estimation of $n_m$ due to the harmonic assumption of the channel's confinement, and there may be additional sources of resistance in series with the channel that limit $I_N(0)$ at the smallest $\nu_x$~\cite{fabritius_irreversible_2023}.

The behavior of $I_N(0)$ can be qualitatively understood following simple arguments based on Pauli's exclusion and Heisenberg's uncertainty principles~\cite{martin_wave-packet_1992, batra_origin_1998, albert_electron_2012}. These yield the result that the particle current flowing between two fermionic reservoirs is given by the energy bandwidth available to form fermionic wave-packets in the channel divided by Planck's constant $I_N = \Delta E/h$. Essentially, access to a larger region in phase space $\Delta E$ allows for denser packing of fermions in real space and therefore a larger current. This argument rooted in the energy-time domain is similar to the explanation of how hydrodynamic flows of electrons more efficiently occupy phase space to suppress the Landauer-Sharvin resistance~\cite{kumar_imaging_2022}. Without interactions, the only contribution to $\Delta E$ is the chemical potential bias $\Delta\mu$, which yields the current scale $I_F(0)$ and conductance quantum $1/h$~\cite{krinner_observation_2014}. Additional energy scales, for example the pairing energy $\Delta$, can enter $\Delta E$ via mechanisms such as a beyond-mean-field generalization of MAR~\cite{choi_andreev_2000, aggarwal_unconventional_2016}, as spectral broadening due to strong interactions~\cite{haussmann_spectral_2009} allows for MAR processes that are forbidden in mean-field theories.

In the bottom row of Fig.~\ref{fig:fig3} (c,f,i), we extend our central observation first presented in Fig.~\ref{fig:fig1} on the universality of the entropy advectively transported per particle, aka Seebeck coefficient $\alpha_c$. We see that, within measurement uncertainty, $\alpha_c$ is independent of the channel properties $V_g$ and $\nu_x$ and depends only on the reservoir properties $1/k_F a$ and $S/N k_B$, increasing monotonically with both. The imperfect collapse of $\alpha_c$ in panels (f) and (i) may be due to the fluctuations in the prepared $S/Nk_B$ (App.~\ref{app:thermodynamics}). This independence of channel detail is a clear departure from the Mott-Cutler formula for Fermi liquids, which predicts that $\alpha_c$ depends strongly on the geometry and density in the channel via its energy-dependent transmission and inter-particle scattering rate~\cite{brantut_thermoelectric_2013, hausler_interaction-assisted_2021, grenier_thermoelectric_2016, gourgout_seebeck_2022}. We found in previous work at unitarity with reservoirs in the normal phase, channel in the 2D-3D crossover, and smaller $\abs{V_g}$ that $\alpha_c$ varied with $V_g$ in qualitative consistency with the Mott-Cutler formula~\cite{hausler_interaction-assisted_2021}. While the particle-hole asymmetry of the reservoirs' fermionic spectrum increases when tuning from the BCS to BEC side~\cite{haussmann_spectral_2009}, it is not clear \textit{a priori} if this influences $\alpha_c$ in the regime of strongly-coupled reservoirs and nonlinear response as it does in the tunnel junction and linear response limits~\cite{kruchkov_thermoelectric_2020}. An alternative picture may be that increasing $1/k_F a$ not only leads to stronger pairing but also to larger fluctuations in the pairing field~\cite{zwerger_bcs-bec_2012} whose mean value yields the current scale $I_N(0)$ and fluctuations yield the average entropy transported per particle $\alpha_c$. A seemingly universal relationship between the Seebeck coefficient and thermodynamic properties has also been observed in strongly correlated electronic systems in the linear response regime~\cite{behnia_thermoelectricity_2004}.

\begin{figure}
    \centering
    \includegraphics{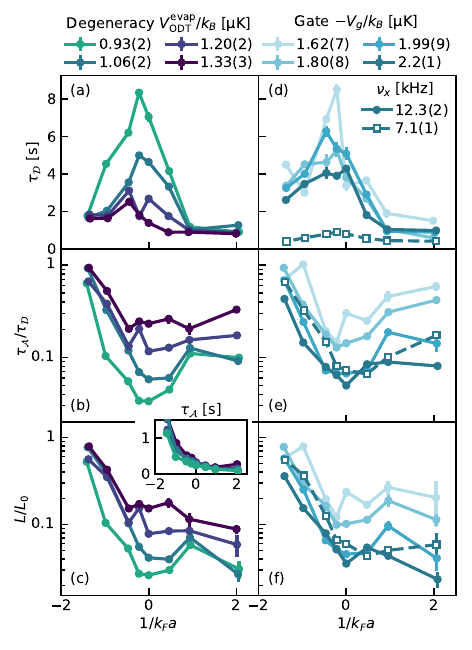}
    \caption{\textbf{Breakdown of the Wiedemann-Franz law at unitarity and its restoration in the BCS limit}. The exponential timescale of entropy diffusion $\tau_\mathcal{D}$ is maximum around unitarity and falls off on either side of the Feshbach resonance for all degeneracies (a) and gate potentials (d), in both 1D [$\nu_x=\SI{12.3(2)}{kHz}$] and the 1D-2D crossover [$\nu_x=\SI{7.1(1)}{kHz}$]. Although the advective timescale $\tau_\mathcal{A}$ monotonically decreases from BCS to BEC, the ratio $\tau_\mathcal{A}/\tau_\mathcal{D}$ (b,e) has a robust minimum around unitary. The Lorenz number $L$ estimated from $\tau_\mathcal{A}/\tau_\mathcal{D}$ (see text) and normalized to the Wiedemann-Franz value $L_0=\pi^2k_B^2/3$ (c,f) is strongly suppressed at unitarity but approaches unity in the BCS limit. The teal data in (a) and (d) both have $V_\mathrm{ODT}^\mathrm{evap}/k_B = \SI{1.06(2)}{\micro K}$ and $-V_g/k_B = \SI{2.2(1)}{\micro K}$. The advective timescales $\tau_\mathcal{A}$ used in (e,f) come from the data in Fig.~\ref{fig:fig3}(d) with $V_\mathrm{ODT}^\mathrm{evap}/k_B = \SI{0.93(2)}{\micro K}$, though this lower $V_\mathrm{ODT}^\mathrm{evap}$ does not significantly affect $\tau_\mathcal{A}$ as shown in the inset of (c).}
    \label{fig:fig4}
\end{figure}

\section{Diffusively limited dynamics}
In Fig.~\ref{fig:fig4}, we study the characteristics of the diffusively limited eigenmode in the 1D channel [$\nu_x=\SI{12.3(2)}{kHz}$] as functions of $1/k_F a$, $V_\mathrm{ODT}^\mathrm{evap}$ (a-c), and $V_g$ (d-f). The data in Fig.~\ref{fig:fig2}(d-f) correspond to $V_\mathrm{ODT}^\mathrm{evap}/k_B=\SI{0.93(2)}{\micro K}$ in Fig.~\ref{fig:fig4}(a-c). The primary characteristics of this eigenmode are the exponential timescale $\tau_\mathcal{D} \propto G_T^{-1}$ and its ratio to the effective advective timescale
$\tau_\mathcal{A}/\tau_\mathcal{D} \propto L$, where $\tau_\mathcal{A}\approx\Delta N(0)/2I_N(0)$ (see App.~\ref{app:eigenmodes_of_transport} for details). For varying degeneracy $V_\mathrm{ODT}^\mathrm{evap}$ [Fig.~\ref{fig:fig4}(a-c)], we obtain $\tau_\mathcal{A}$ from data shown in Fig.~\ref{fig:fig3}(a-c) taken under the same conditions. For varying $V_g$ [Fig.~\ref{fig:fig4}(d-f)], we obtain $\tau_\mathcal{A}$ from Fig.~\ref{fig:fig3}(d-f) taken at 10\% lower $V_\mathrm{ODT}^\mathrm{evap}$, though the change in $\tau_\mathcal{A}$ with $V_\mathrm{ODT}^\mathrm{evap}$ is negligible as shown in the inset of Fig.~\ref{fig:fig4}(c). Fig.~\ref{fig:fig4}(a,d) show that $\tau_\mathcal{D}$ is peaked, i.e. entropy diffusion is slowest, around unitarity for all $V_\mathrm{ODT}^\mathrm{evap}$ and $V_g$ and reaches similar values on either side of the resonance. Diffusion becomes slower with decreasing $S/Nk_B$ for all interactions, which can be understood as lower temperature reducing the density of heat-carrying excitations~\cite{smith_transport_1989}. On the other hand, diffusion becomes faster with increasing $\abs{V_g}$, which may result from an increasing number of occupied transmission modes $n_m$. We estimate this effect to decrease $\tau_\mathcal{D}$ by a factor of $\approx10$ over this range of $V_g$ (App.~\ref{app:thermodynamics}) while we observe only a factor of $\approx2$. This discrepancy may be due to $V_g$ enhancing the local pairing gap $\Delta$ which suppresses quasiparticle-induced diffusion~\cite{pershoguba_thermopower_2019}. It would follow that entropy carriers other than fermionic quasiparticles, for example collective modes~\cite{uchino_role_2020} or incoherent pairs~\cite{uchino_bosonic_2020}, become responsible for diffusion on the BEC side where $\Delta$ continues to increase (Fig.~\ref{fig:fig3_supplement}).

The ratio of the advective and diffusive timescales $\tau_\mathcal{A}/\tau_\mathcal{D} \approx L \ell_r/[\ell_r + (\alpha_c-\alpha_r)^2]^2$, being proportional to the Lorenz number $L$, quantifies the relative strength of the two eigenmodes without precise knowledge of the reservoir response coefficients $\alpha_r$ and $\ell_r$. This ratio, plotted in Fig.~\ref{fig:fig4}(b,e), decreases with lower $S/N k_B$ and larger $\abs{V_g}$, but the location of the minimum around unitarity is robust for all $S/N k_B$ and $\abs{V_g}$. Using the values of $\alpha_r$ and $\ell_r$ estimated from the equation of state in a harmonic potential across the BCS-BEC crossover (App.~\ref{app:thermodynamics}), we plot in Fig.~\ref{fig:fig4}(c,f) the extracted $L$ normalized by the Wiedemann-Franz value $L_0=\pi^2k_B^2/3$. Although the real reservoir coefficients might deviate from the harmonic estimates~\cite{fabritius_irreversible_2023}, the Wiedemann-Franz law is clearly violated at unitarity~\cite{husmann_breakdown_2018, fabritius_irreversible_2023} and restored in the BCS limit. Moreover, in the 1D-2D crossover regime of the channel where the diffusion time is an order of magnitude faster [$\nu_x=\SI{7.1(1)}{kHz}$, open squares in Fig.~\ref{fig:fig4}(d,e,f)], we observe the same behavior of $\tau_\mathcal{A}/\tau_\mathcal{D}$ and $L$ vs.~$1/k_Fa$. This shows that the violation of the Wiedemann-Franz law at unitarity as well as its restoration in the BCS limit is not due to the details of the channel.

\section{Conclusions}
We have studied the coupled transport of particles and entropy over a large range of parameters corresponding to both the thermodynamic properties of the reservoirs in the BCS-BEC crossover and the potential energy landscape in the channel. We have found that the average entropy advectively transported per particle (the Seebeck coefficient), a property of the far-from-equilibrium currents in the channel, depends not on the details of the channel but only on the universal equilibrium properties of the reservoirs. Moreover, we find that the nonlinear phenomenological model of thermoelectric transport we originally developed in the superfluid phase at unitarity~\cite{fabritius_irreversible_2023} describes the system over the full parameter range explored here, smoothly connecting regimes of linear and nonlinear response. Even though the assumption of local equilibrium in the contacts underestimates $\alpha_c$ by orders of magnitude~\cite{fabritius_irreversible_2023}, it partially reproduces the observed nonlinear current scale with the estimated local superfluid gap $\Delta$ and number of occupied transmission modes $n_m$. We observed that entropy diffusion is slowest around unitarity, where the Wiedemann-Franz law is most egregiously violated, suggesting a change in the nature of the excitations responsible for entropy diffusion when crossing from the BCS to BEC sides. These observations call for microscopic explanations of this emergent universality beyond the regime of linear hydrodynamics.

\section*{Acknowledgments}
We are grateful to Alexander Frank for his contributions to the electronics of the experiment. We thank Rudolf Haussmann for providing us the results of his calculations of thermodynamic properties across the BCS-BEC crossover. We are grateful to Alex G\'omez Salvador and Eugene Demler for their collaboration. We thank Wilhelm Zwerger for his valuable feedback on the manuscript. We acknowledge the Swiss National Science Foundation (Grants No.~212168, UeM019-5.1, 182650, and TMAG-2~209376) and European Research Council advanced grant TransQ (Grant No.~742579) for funding.

\appendix
\section{Experimental setup and sequence}
\label{app:experimental_setup_and_sequence}

\begin{figure}
    \centering
    \includegraphics[width=\hsize]{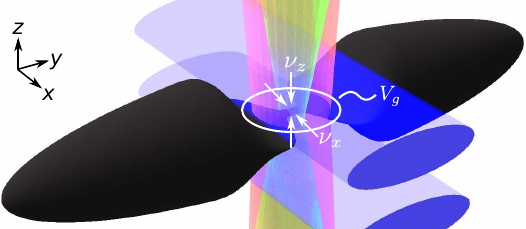}
    \caption{\textbf{Optically defining the reservoirs and tunable ballistic channel.} The isopotential surface (black) shows the reservoirs and channel. The blue $\mathrm{TEM}_{01}$ beam generates the vertical transverse confinement $\nu_z$, the green $\mathrm{TEM}_{01}$ beam generates the horizontal transverse confinement $\nu_x$, and the red gaussian beam generates the gate potential $V_g$. The optical dipole and magnetic traps are not shown but are incorporated into the isopotential surface.}
    \label{fig:setup_supplement}
\end{figure}

In Fig.~\ref{fig:setup_supplement} we show the system during transport. The gas is a balanced mixture of the first and third lowest energy spin states. It is trapped by a combination of an optical dipole trap (ODT) propagating along $y$ and the saddle potential of the magnetic field $B$ that addresses the Feshbach resonance at $B=\SI{690}{G}$, which primarily confines along $y$. We separate the cigar-shaped cloud into two reservoirs connected by a channel (shown in black) by intersecting two repulsive $\mathrm{TEM}_{01}$-like beams at the center (blue and green beams). The confinement frequencies of the beams vary along the direction of transport $y$ and reach the values $\nu_x$ and $\nu_z$ at the center where the beam intensities are maximum. An attractive gaussian beam propagating along $z$ (red), which has approximately the same waist as the blue beam along the direction of transport, produces a gate potential that varies with the intensity and reaches the value $V_g$ at the center of the channel. An additional repulsive gaussian beam (not shown) that is nearly the same size as the green beam acts as a wall that we use to separate the reservoirs and to block and allow transport by switching it on and off. See the supplement of Ref.~\cite{fabritius_irreversible_2023} for further details of the beams and the potential energy landscape.

\begin{figure*}
    \centering
    \includegraphics{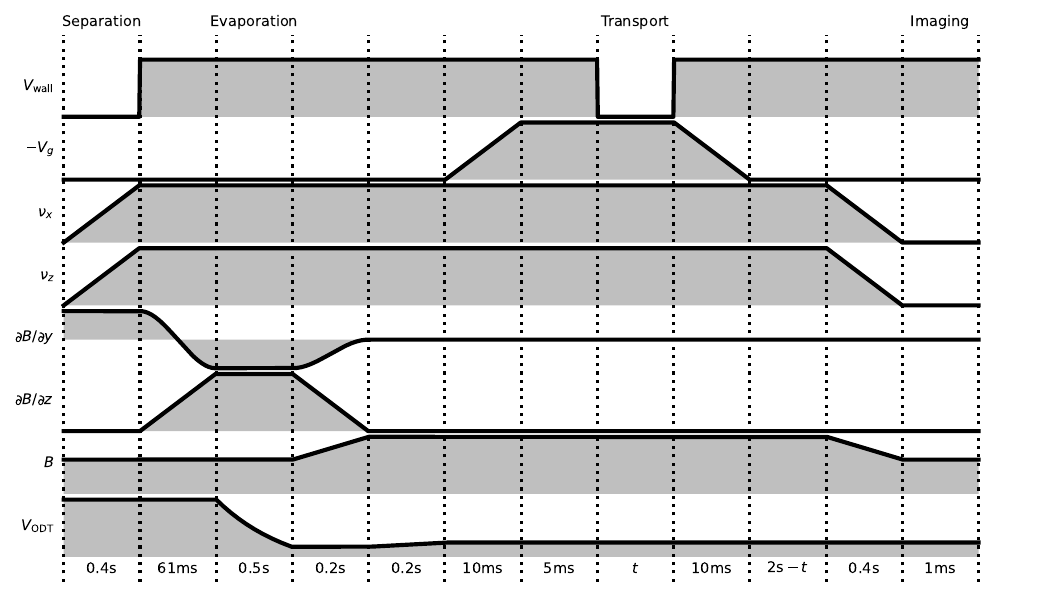}
    \caption{\textbf{Experimental sequence with the duration of each step.} The bulk cloud is separated into two reservoirs connected by the channel by ramping up the power of the channel beams, which set the confinement frequencies $\nu_x, \nu_z$ during transport, and subsequently switching on the wall. Forced optical evaporation is controlled with the potential depth of the dipole trap $V_\mathrm{ODT}$ and the magnetic field gradient along gravity $\pdv*{B}{z}$ to further reduce the trap depth. Following evaporation, $V_\mathrm{ODT}$ is slightly increased to avoid further evaporation. The magnetic field $B$ sets the interaction strength, which is unitary during evaporation and imaging but possibly elsewhere on the crossover during transport. The center position of the trap with respect to the channel is set with the magnetic field gradient along the transport direction $\pdv*{B}{y}$. We can prepare an almost arbitrary combination of $\Delta N(0)$ and $\Delta S(0)$ with the value of $\pdv*{B}{y}$ during separation and evaporation, while we always align the trap center to the channel during transport to ensure that $\Delta N=\Delta S=0$ at equilibrium $\Delta\mu=\Delta T=0$. Transport is initiated by ramping up the gate potential $V_g$ and switching off the wall, and halted by switching the wall back on and ramping $V_g$ back down.}
    \label{fig:sequence}
\end{figure*}

In Fig.~\ref{fig:sequence} we show the relevant section of the experimental sequence. We begin with a non-degenerate ($T/T_F\sim1$) gas at unitarity in the harmonic trap of the ODT and magnetic saddle potential. We prepare the imbalances $\Delta N$ and $\Delta S$ with the position of the trap center along the direction of transport $y$, set with a magnetic field gradient $\pdv*{B}{y}$ generated by an additional set of coils, during the separation and evaporation steps. With the gradient set to $(\pdv*{B}{y})_\mathrm{sep}$, we ramp up the power of the two $\mathrm{TEM}_{01}$ beams that generate the confinements $\nu_x$ and $\nu_z$ and switch on the wall at the end of the ramps to separate the cloud into two reservoirs. We then ramp the gradient to $(\pdv*{B}{y})_\mathrm{evap}$ to compress one reservoir and decompress the other such that, during the forced evaporation induced by a field gradient along gravity $\pdv*{B}{z}$ and a ramp down of the dipole trap depth to $V_\mathrm{ODT}=V_\mathrm{ODT}^\mathrm{evap}$, the evaporation efficiency of the two reservoirs are tuned independently. Operationally, we prepare the advectively dominated eigenmode primarily with a large $(\pdv*{B}{y})_\mathrm{sep}$ to set a particle imbalance $\Delta N$ and fine tune the entropy imbalance $\Delta S$ with $(\pdv*{B}{y})_\mathrm{evap}$ such that the trajectory in state space is linear. Conversely, we prepare the diffusively limited eigenmode primarily with a large $(\pdv*{B}{y})_\mathrm{evap}$ to prepare $\Delta S$ and fine tune $\Delta N$ with $(\pdv*{B}{y})_\mathrm{sep}$ to ensure a linear trajectory. Because the advective mode is generally much faster than the diffusive mode, the preparation of $\Delta S(0)$ need not be so precise to probe the advectively dominated eigenmode, so we prepare the same $\Delta N(0), \Delta S(0)$ -- i.e.~the same $(\pdv*{B}{y})_\mathrm{sep}$ and $(\pdv*{B}{y})_\mathrm{evap}$ -- for all interaction strengths in Fig.~\ref{fig:fig2}(a-c). However, the diffusively limited eigenmode requires a different $(\pdv*{B}{y})_\mathrm{sep}$ for different degeneracies $S/N$. We exclude some data points from the fits since this preparation was not perfect for different interactions and therefore still contains some of the advectively dominated eigenmode.

After evaporation, we adiabatically ramp the Feshbach field $B$ to a desired interaction strength $1/k_F a$ such that $N(0)$, $S(0)$, $\Delta N(0)$, and $\Delta S(0)$ are all conserved. Due to the saddle potential of the Feshbach field, the geometric mean of the harmonic trap frequencies in all directions $\bar{\omega}/2\pi$ is 95.7(6) Hz at unitarity and varies by $\pm2\%$ over the range of $a$ we explore. Simultaneously, we ramp $\pdv*{B}{y}$ such that the trap center is aligned to the channel to ensure that the imbalances vanish $\Delta N = \Delta S = 0$ at equilibrium $\Delta\mu = \Delta T = 0$. We also ramp up $V_\mathrm{ODT}$ by a small amount to ensure no further evaporation occurs. Next, we ramp up the gate potential and switch off the wall for a time $t$ to allow transport before switching back on the wall and ramping down the gate. We then wait for a time $\SI{2}{s}-t$ such that the total duration of the time spent at each interaction strength is kept constant at \SI{2.225}{s} independently of $t$ to preserve thermalization of the experimental apparatus. We have confirmed that there is no systematic drift in the alignment of the high-resolution projection optics with the interaction strength, though some small fluctuations were present which have the effect of changing the potential energy in the channel. Finally, we adiabatically ramp down the channel beam powers and the magnetic field back to unitarity and take absorption images of both spin states in both reservoirs to extract $N(t)$, $S(t)$, $\Delta N(t)$, and $\Delta S(t)$ using standard procedures described in the supplement of Ref.~\cite{fabritius_irreversible_2023}.

\section{Eigenmodes of transport}
\label{app:eigenmodes_of_transport}

The phenomenological model in Eq.~\ref{eq:phenomenological_model} describes the intertwined dynamics of $\Delta N$ and $\Delta S$ over the full parameter range we explore here. In the linear response regime $\sigma \gg \Delta\mu + \alpha_c \Delta T$, the hyperbolic tangent function tanh can be linearized and Eq.~\ref{eq:phenomenological_model} reduces to the usual linear response model of thermoelectric transport~\cite{husmann_breakdown_2018} that yields exponential relaxation captured by the matrix $\vb*{\Lambda}$
\begin{align}
    \label{eq:linear_eom}
    \dv{t} \pmqty{\Delta N \\ \Delta S} &= -\frac{4G}{\kappa} \pmqty{1 & \alpha_c \\ \alpha_c & L + \alpha_c^2} \pmqty{1 & \alpha_r \\ \alpha_r & \ell_r + \alpha_r^2}^{-1} \pmqty{\Delta N \\ \Delta S} \nonumber \\
    &\equiv -\vb{\Lambda} \pmqty{\Delta N \\ \Delta S}.
\end{align}
This matrix has two pairs of eigenvalues $\tau_\pm^{-1}$ and eigenvectors $\vb{v}_\pm = (1, \Pi_\pm)^T$ comprising its eigenmodes $\mathcal{E}_\pm$ that are generally hybridizations of the bare advective and diffusive modes
\begin{align}
    \label{eq:eigensystem}
    \begin{split}
    \tau_\pm^{-1} &= \frac{L + \ell_r + (\alpha_c-\alpha_r)^2 \pm \lambda}{2\ell_r} \tau_0^{-1} \\
    \Pi_\pm &= \frac{L - \ell_r + \alpha_c^2 - \alpha_r^2 \pm \lambda}{2(\alpha_c-\alpha_r)} \\
    \lambda &= \sqrt{[L + \ell_r + (\alpha_c-\alpha_r)^2]^2 - 4 L \ell_r}
    \end{split}
\end{align}
where $\tau_0 = \kappa/4G$ is the natural timescale of evolution. Our observation of purely irreversible, damped dynamics (i.e. no oscillations or exponential growth) imposes that $\lambda$ is a real number and $\lambda \geq 0$, with $\lambda=0$ only at $L=\ell_r$ and $\alpha_c=\alpha_r$. This also implies that $\tau_+ \leq \tau_-$. When the system traces a straight line through state space with a slope $\Pi_\pm = \dv*{\Delta S}{\Delta N}$, only one of these two components evolves significantly during the observation; in other words, the system traces a path parallel to only one of the two eigenvectors. The observation that the faster eigenmode exhibits non-exponential relaxation in some regime allows us to identify it as being dominated by the nonlinear advective mode $\mathcal{A} = \mathcal{E}_+$, while the slower eigenmode, which always exhibits exponential relaxation, is limited by the linear diffusive mode $\mathcal{D} = \mathcal{E}_-$. As the system's response smoothly crosses over from linear to nonlinear response, $\mathcal{A}$ becomes identical to the bare advective mode.

Given the constraint that the path in state space is a line $\Delta S(t) = \Pi_\mathcal{E} \Delta N(t) + \Delta S_\mathrm{neq}$, we can rewrite Eq.~\ref{eq:phenomenological_model} as
\begin{equation}
    I_N(\Delta N) = I_\mathrm{exc} \tanh\bqty{\frac{\Delta N - \Delta N_\mathcal{E}(\infty)}{\sigma_\mathcal{E}}}
\end{equation}
with the re-scaled parameters
\begin{align*}
    \sigma_\mathcal{E} &= \frac{\kappa\ell_r}{2[\ell_r + (\alpha_r-\alpha_c)(\alpha_r-\Pi_\mathcal{E})]} \sigma \\
    \Delta N_\mathcal{E}(\infty) &= \frac{\alpha_r - \alpha_c}{\ell_r + (\alpha_r-\alpha_c)(\alpha_r-\Pi_\mathcal{E})} \Delta S_\mathrm{neq}.
\end{align*}
The equation of motion $\dv*{\Delta N}{t} = -2 I_N$ then has the analytic solution given in Eq.~\ref{eq:analytic_nonlinear}. Re-writing Eq.~\ref{eq:phenomenological_model} as Eq.~\ref{eq:analytic_nonlinear} is beneficial because it reduces the number of parameters needed to characterize the system and can be directly fitted to our data without explicitly introducing the thermodynamic response functions of the reservoirs. In the nonlinear regime $\sigma_\mathcal{E} \ll \Delta N(0) - \Delta N_\mathcal{E}(\infty)$, Eq.~\ref{eq:analytic_nonlinear} simplifies to a piecewise linear function 
\begin{equation}
    \label{eq:analytic_kinky}
    \Delta N(t) \approx \begin{cases}
        \Delta N(0) - 2 I_\mathrm{exc}t & t < \dfrac{\Delta N(0) - \Delta N_\mathcal{E}(\infty)}{2I_\mathrm{exc}} \\
        \\
        \Delta N_\mathcal{E}(\infty)  & t \geq \dfrac{\Delta N(0) - \Delta N_\mathcal{E}(\infty)}{2I_\mathrm{exc}}
    \end{cases}
\end{equation}
while in the linear regime $\sigma_\mathcal{E} \gg \Delta N(0) - \Delta N_\mathcal{E}(\infty)$, the relaxation becomes exponential
\begin{equation}
    \label{eq:analytic_linear}
    \Delta N(t) \approx [\Delta N(0) - \Delta N_\mathcal{E}(\infty)] e^{-2I_\mathrm{exc}t/\sigma_\mathcal{E}} + \Delta N_\mathcal{E}(\infty).
\end{equation}
In this linear response limit, $I_\mathrm{exc}$ and $\sigma_\mathcal{E}$ are ill-defined and only their ratio, which defines the exponential timescale $\tau_\mathcal{E}^{-1} = 2I_\mathrm{exc}/\sigma_\mathcal{E}$, is physically meaningful. It is therefore advantageous to characterize the system with quantities that are well-defined for both exponential and non-exponential evolution, for which we choose the initial current $I_N(0)$. This is straightforward to extract directly from the fitted $\Delta N(t)$ and simplifies to $I_\mathrm{exc}$ in the nonlinear regime and $[\Delta N(0) - \Delta N_\mathcal{E}(\infty)]/2\tau_\mathcal{E}$ in the linear regime. The term $\Delta N_\mathcal{E}(\infty)$ that would be finite in a non-equilibrium steady state~\cite{husmann_breakdown_2018, fabritius_irreversible_2023} is not well-constrained in our fits when the relaxation time is much longer than the experiment. However, we prepare the system such that $\Delta N_\mathcal{E}(\infty)$ is much smaller than $\Delta N_\mathcal{E}(0)$. We therefore extract the exponential timescale in the linear response regime and the effective timescale in the nonlinear regime using the approximation $\tau_\mathcal{E} \approx \Delta N(0)/2I_N(0)$; this expression was used to obtain $\tau_\mathcal{A}$ and $\tau_\mathcal{D}$ in Fig.~\ref{fig:fig4}. In the inset between panels (a) and (b) in Fig.~\ref{fig:fig3} that plots $\sigma_\mathcal{A}$, the data sets are considered linear and therefore excluded from the plot when any of the following criteria are fulfilled: the fitted $\sigma_\mathcal{A}$ is larger than $\Delta N(0)$, the standard error of the fitted $\sigma_\mathcal{A}$ is larger than $N$, or the fitted initial and final imbalances satisfy $\Delta N(\infty) > \Delta N(0)/5$ in which case the system did not relax enough to sufficiently constrain $\Delta N(\infty)$ or $\sigma_\mathcal{A}$.

The observation of extremely slow diffusively limited relaxation and the related non-equilibrium steady state~\cite{husmann_breakdown_2018, fabritius_irreversible_2023} implies that $\tau_\mathcal{D} \rightarrow \infty$ and therefore $L \ll \ell_r + (\alpha_c-\alpha_r)^2$. The nonlinearity of the advective mode further implies that $L = G_T/TG \rightarrow 0$ since $G = I_\mathrm{exc}/\sigma$ diverges. We therefore Taylor expand $\tau_\mathcal{E}^{-1}$ and $\Pi_\mathcal{E}$ about $L=0$ to obtain
\begin{align}\begin{split}
    \label{eq:expansion_small_L}
    \tau_\mathcal{A}^{-1} &\approx \bqty{1 + \frac{(\alpha_c-\alpha_r)^2}{\ell_r}} \qty{1 + \frac{L(\alpha_c-\alpha_r)^2}{[\ell_r + (\alpha_c-\alpha_r)^2]^2}} \tau_0^{-1} \\
    \tau_\mathcal{D}^{-1} &\approx \frac{L}{\ell_r + (\alpha_c-\alpha_r)^2} \tau_0^{-1}
    = \frac{4G_T}{T\kappa[\ell_r + (\alpha_c-\alpha_r)^2]} \\
    \Pi_\mathcal{A} &\approx \alpha_c + \frac{L(\alpha_c-\alpha_r)}{\ell_r + (\alpha_c-\alpha_r)^2} \\
    \Pi_\mathcal{D} &\approx \alpha_r + \frac{\ell_r}{\alpha_r - \alpha_c} \bqty{1 - \frac{L}{\ell_r + (\alpha_r-\alpha_c)^2}}
\end{split}\end{align}
The slope of the path that $\mathcal{A}$ traces through state space therefore directly yields the entropy advectively carried per particle $\Pi_\mathcal{A} \approx \alpha_c$; this approximation is exact for non-exponential relaxation. The advective timescale is inversely proportional to the conductance $\tau_\mathcal{A}^{-1} \propto G$, the diffusive timescale is inversely proportional to the thermal conductance $\tau_\mathcal{D}^{-1} \propto G_T$, and the ratio of the two is proportional to the Lorenz number $\tau_\mathcal{A}/\tau_\mathcal{D} \approx L \ell_r/[\ell_r + (\alpha_c-\alpha_r)^2]^2$.

The limit discussed above $L\rightarrow0$ leads to the largest separation of timescales possible, i.e. $\lambda$ tends to its maximum value $\ell_r + (\alpha_c-\alpha_r)^2$. We now consider the opposite limit where $\lambda\rightarrow0$ and the two eigenmodes become degenerate. As can be seen in the expression for $\lambda$ in Eq.~\ref{eq:eigensystem}, this limit is achieved only in the limit of $\alpha_c=\alpha_r$ and $L=\ell_r$. This limit is not well-defined and depends on the direction from which the point $\alpha_r, \ell_r$ in $\alpha_c, L$ space is approached, though the non-singular part of the limit is $\Pi_\mathcal{A} \approx (\alpha_c+\alpha_r)/2 - (3L + \ell_r)/4\sqrt{L} \rightarrow \alpha_c - \sqrt{L}$.

\section{Natural scales of transport}
\label{app:natural_scales_of_transport}

We wish to compare the observables $I_N(0)$ and $\tau_\mathcal{E}$ to some characteristic scales of the system to properly non-dimensionalize these quantities. The usual strategy in the context of the BCS-BEC crossover is to compare them to the relevant Fermi scale, defined as the value of any quantity for a non-interacting Fermi gas at zero temperature. In this condition, response is linear and the Seebeck coefficient $\alpha_c$, thermal conductance $G_T$, and entropy bias $\Delta S$ are all zero~\cite{grenier_thermoelectric_2016}, so the particle current is simply $I_N = G\Delta\mu = 2G\Delta N/\kappa$. At $T=0$ where the chemical potential equals the Fermi energy -- approximated by $E_F = \hbar^2k_F^2/2m = \hbar\bar{\omega}(3N)^{1/3}$ when neglecting the anharmonicity of the channel -- the compressibility is $\kappa = \pdv*{N}{\mu} = \pdv*{N}{E_F} = 3N/E_F$. Furthermore, the quantum of conductance including spin degeneracy is $G=2/h$. Therefore, the Fermi scale for the initial current is $I_F(0) = 4E_F\Delta N(0)/3hN$. We use the same approach to define the Fermi scale for the exponential relaxation $\tau_F$ defined by $\dv*{\Delta N}{t} = -\Delta N/\tau_F$ to find $\tau_F = \Delta N(0)/2I_F(0) = 3hN/8E_F$. In the limit of zero temperature, $\alpha_c = \alpha_r = 0$ while $L = \ell_r = L_0$ for a non-interacting Fermi gas, so $\tau_\mathcal{A}$ and $\tau_\mathcal{D}$ both simplify to $\tau_F$.

In additional to the Fermi scales that are applicable within linear response, we compare the observed current to the nonlinear current scale derived from microscopic models based on multiple Andreev reflection (MAR)~\cite{husmann_connecting_2015, huang_superfluid_2023, visuri_dc_2023}. These models predict an initial current given by $I_\mathrm{MAR}(0) = n_m[16\Delta/3h + I_F(0)]$ where $n_m$ is the number of occupied transverse modes in the channel in equilibrium and $\Delta$ is the local equilibrium superfluid gap at the contacts to the channel as in previous work~\cite{huang_superfluid_2023, fabritius_irreversible_2023}. See Fig.~\ref{fig:fig3_supplement} for estimates of these quantities. The excess current $I_\mathrm{exc}$ is explicitly given by the superfluid-induced term $\propto\Delta$ and the Fermi term $\propto I_F(0)$ constitutes a normal, Ohmic component. We find that substituting the contribution $I_F(0)$ with the more accurate value $\Delta\mu(0)/h$ computed from $\Delta N(0)$, $\Delta S(0)$, and the thermodynamic response functions does not improve the data collapse nor independence of $I_N(0)/I_\mathrm{MAR}(0)$ on $1/k_F a$ in Fig.~\ref{fig:fig3}(b,e,h) since the superfluid term $16\Delta/3h$ dominates for all data sets.

\section{Thermodynamics of the reservoirs over the BCS-BEC crossover}
\label{app:thermodynamics}

\begin{figure*}
    \centering
    \includegraphics{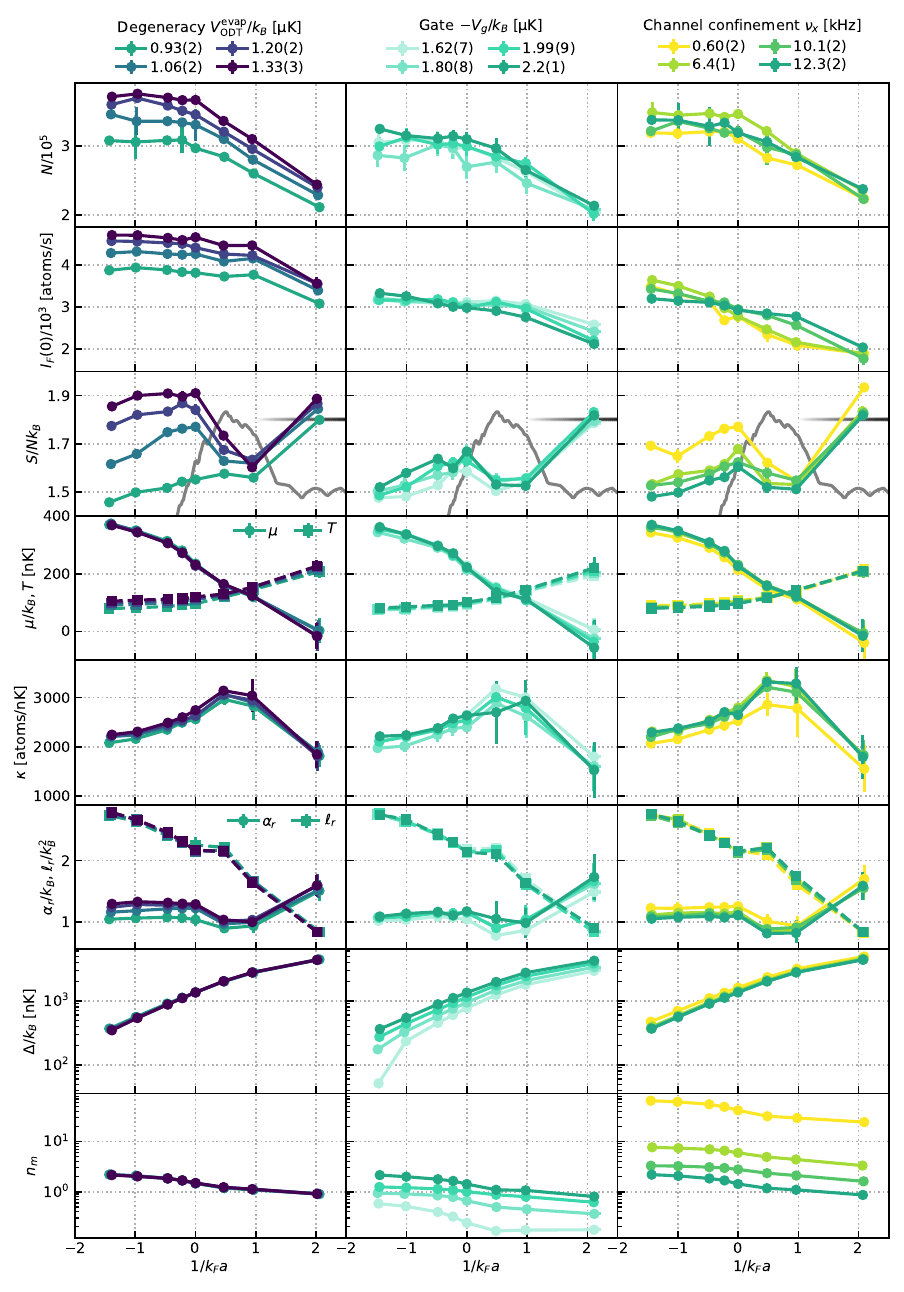}
    \caption{Thermodynamic properties of the system from Fig.~\ref{fig:fig3} computed from the measured $N$ and $S/N k_B$.}
    \label{fig:fig3_supplement}
\end{figure*}

To estimate the thermodynamic properties of the reservoirs as functions of $S/Nk_B$ and $1/k_Fa$, we use the calculations for the homogeneous gas~\cite{haussmann_thermodynamics_2007} and apply the local density approximation~\cite{haussmann_thermodynamics_2008} with the harmonic trapping potential using the known average trap frequency $\bar{\omega}$. From the calculations, we have the local property expressed in a universal form, for example the particle density $n(\mu,T,a) = \lambda_T^{-3} f_n(\mu/k_B T, \lambda_T/a)$, entropy density $s(\mu,T,a) = k_B\lambda_T^{-3} f_s(\mu/k_B T, \lambda_T/a)$, and superfluid gap $\Delta(\mu,T,a) = k_B T f_\Delta(\mu/k_B T, \lambda_T/a)$ where $\lambda_T = \sqrt{2\pi\hbar^2/m k_B T}$ is the thermal de Broglie wavelength. Outside the computed range of $\mu/k_B T$ where the system is in the normal phase, we use the analytical result for a weakly interacting fermi gas in the BCS limit $\lambda_T/a < -1$~\cite{su_low-temperature_2003} and the second order virial expansion for $\lambda_T/a \geq -1$~\cite{jager_precise_2023}. Because the numerical methods erroneously predict a first order phase transition and therefore multivalued thermodynamic functions, we use the same approach as in Ref.~\cite{haussmann_thermodynamics_2008} to correct for this and connect the normal and superfluid branches with a kink where they are closest. This discontinuity is the origin of the noise in the critical entropy per particle $(S/Nk_B)_c$ in the inset between panels (b) and (c) in Fig.~\ref{fig:fig3}.

The trap-averaged quantities such as the total atom number $N$ and entropy $S$ are spatial integrals of these functions at constant $T$ and $a$ but spatially-varying local chemical potential $\breve{\mu}(\vb{r}) = \mu - (m/2)\bar{\omega}^2r^2$, i.e. $N(\mu,T,a,\bar{\omega}) = \int n[\breve{\mu}(\vb{r}), T, a] \dd[3]{r} = (k_B T/\hbar\bar{\omega})^3 f_N(\mu/k_B T, \lambda_T/a)$ and $S(\mu,T,a,\bar{\omega}) = \int s[\breve{\mu}(\vb{r}), T, a] \dd[3]{r} = k_B (k_B T/\hbar\bar{\omega})^3 f_S(\mu/k_B T, \lambda_T/a)$, and the thermodynamic response functions $\kappa = (\pdv*{N}{\mu})_T$, $\alpha_r = -(\pdv*{N}{T})_\mu/\kappa = (\pdv*{S}{N})_T$, and $\ell_r = (\pdv*{S}{T})_\mu/\kappa - \alpha_r^2 = (\pdv*{S}{T})_N/\kappa$ are derivatives thereof~\cite{fabritius_irreversible_2023}. We note that, in our previous work at unitarity~\cite{fabritius_irreversible_2023} wherein we employed the measured equation of state~\cite{ku_revealing_2012}, the fitted $\alpha_r$ approximately matched the computed value in the anharmonic trap, but the fitted $\ell_r$ was significantly smaller than its estimate, possibly due to the complex potential energy landscape of the anharmonic trap.

It is straightforward to show that $S/Nk_B$ and $1/k_F a$ as defined in the main text are universal functions of $\mu/k_B T$ and $\lambda_T/a$ and do not depend on absolute scales such as $\bar{\omega}$ or $N$. We determine $\mu$ and $T$ of the reservoirs at each $a$ from the known $\bar{\omega}$ and measured $N$ and $S$. The local properties of the channel such as $\Delta$ and $n_m$ are described in the supplement of Ref.~\cite{fabritius_irreversible_2023}: $\Delta$ is computed at the most degenerate point in space at the contacts to the channel, and $n_m$ is the thermally-averaged number of propagating single-particle transmission modes in the channel. The relevant thermodynamic properties of the reservoirs and channel in the runs of Fig.~\ref{fig:fig3} are presented in Fig.~\ref{fig:fig3_supplement}, computed from atom number $N$ and entropy $S$ measured after \SI{2.225}{s} of hold time at each interaction strength but $t=0$ transport time and the approximate the potential energy landscape assuming only harmonic trapping~\cite{fabritius_irreversible_2023}.

\clearpage
\bibliography{references}

\end{document}